\documentclass[aps,dvipsnames,onecolumn,floatfix,preprint,amsmath,amssymb,superscriptaddress]{revtex4-1}
\usepackage{graphicx}%
\relax
\begin{document}

\title{Local magnetic measurements of permanent current paths in
 a natural  graphite crystal}

% Place the author information here.  Please hand-code the contact
% information and notecalls; do *not* use \footnote commands.  Let the
% author contact information appear immediately below the author names
% as shown.  We would also prefer that you don't change the type-size
% settings shown here.

\author{Markus Stiller}
\affiliation{Division of Superconductivity and Magnetism, Felix Bloch Institute
for Solid State Physics, Universit\"{a}t Leipzig,
Linn\'{e}stra{\ss}e 5, D-04103 Leipzig, Germany}
\author{Pablo D. Esquinazi}\email{esquin@physik.uni-leipzig.de}
\affiliation{Division of Superconductivity and Magnetism, Felix Bloch Institute
for Solid State Physics, Universit\"{a}t Leipzig,
Linn\'{e}stra{\ss}e 5, D-04103 Leipzig, Germany}
\author{Christian E. Precker}
\affiliation{Division of Superconductivity and Magnetism, Felix Bloch Institute
for Solid State Physics, Universit\"{a}t Leipzig,
Linn\'{e}stra{\ss}e 5, D-04103 Leipzig, Germany}
\author{Jos\'e Barzola-Quiquia}
\affiliation{Division of Superconductivity and Magnetism, Felix Bloch Institute
for Solid State Physics, Universit\"{a}t Leipzig,
Linn\'{e}stra{\ss}e 5, D-04103 Leipzig, Germany}

% Include the date command, but leave its argument blank.


\begin{abstract}
  A recently reported transition in the electrical resistance of
  different natural graphite samples suggests the existence of
  superconductivity at room temperature. To check whether
  dissipationless electrical currents are responsible for the trapped magnetic
  flux inferred from electrical resistance measurements, we localized
  them using magnetic force microscopy on a natural graphite sample in
  remanent state after applying a magnetic field. The obtained evidence indicates that at room temperature
  a permanent current flows at the
  border of the trapped flux region. The current path vanishes at the same transition temperature
  $T_c \simeq 370~$K as the one obtained from electrical resistance
  measurements on the same sample. The overall results support
  the existence of room-temperature superconductivity at certain
  regions  in the graphite structure and show that the used method
  is suitable to
  localize the superconducting regions.
\end{abstract}

\maketitle
\bigskip

% In setting up this template for *Science* papers, we've used both
% the \section* command and the \paragraph* command for topical
% divisions.  Which you use will of course depend on the type of paper
% you're writing.  Review Articles tend to have displayed headings, for
% which \section* is more appropriate; Research Articles, when they have
% formal topical divisions at all, tend to signal them with bold text
% that runs into the paragraph, for which \paragraph* is the right
% choice.  Either way, use the asterisk (*) modifier, as shown, to

%\section*{Introduction}

To prove the existence of superconductivity in certain small regions
of a macroscopic sample, usual experimental methods that show zero
electrical resistance and/or magnetic flux expulsion are not always
well suitable. This is the case where the superconducting regions
are localized within a two-dimensional interface at which no easy
access for direct electrical contacts to the regions of interest is
possible.  Moreover, if the size of the superconducting regions is
much smaller than the effective London penetration depth, in addition
to demagnetization effects, the flux expulsion, i.e.~the Meissner
effect, might be immeasurable. In this case, an alternative proof for the
existence of superconductivity can rely on the observation of
dissipationless currents that maintain a magnetic flux trapped at
certain regions of the sample or interface.  Recently published
results\cite{pre16} suggest that graphite samples show a
superconducting-like transition at surprisingly high transition temperatures
 $T_c \gtrsim 350~$K. The observation of Bragg peaks in X-ray
diffraction (XRD) measurements corresponding to the two possible stacking orders of
graphite, rhombohedral and hexagonal, suggests their interfaces as the
regions where superconductivity can be localized\cite{mun13,vol13,peo15,hei16} due to
existence of flat bands\cite{fen12,col13,pie15}.

The sample used was cut from a large natural graphite sample from a
mine in Sri Lanka. Before we started the MFM measurements, the sample
was previously characterized with the electrical resistance four-terminals measurements.
The temperature dependence of the resistance and of the remanence
(three point measurements) after applying a certain magnetic field,
agree with published results for similar samples, see~\cite{pre16}.
According to XRD measurements, the sample has two
well defined stacking orders, i.e.~Bernal and rhombohedral
order~\cite{pre16}.  This implies that interfaces between the two
stacking orders exist inside the sample, as TEM pictures
indicate~\cite{pre16,QYRSE08}.  In addition, interfaces between two
crystalline regions of similar stacking order, but which are rotated
around the a common $c$-axis, are possible. The two phases and the
interfaces have large influence on the electrical transport
properties~\cite{ZQSSEKMLL17}.

For the localization of dissipationless currents in a graphite sample produced after we remove an applied magnetic field on the sample
in the zero field cooled (ZFC) state,
we selected an available magnetic force microscope (MFM)
Nanoscope IIIa  from Digital Instruments, which
provides images of a phase signal proportional to the second
derivative of the stray field component, see, e.g., \cite{kir10}. Tapping mode is used to
obtain the topography, the tip is then raised just above the sample
such that a constant separation between sample and tip is
maintained.  The scan height was kept at 200~nm for all
measurements. (local stray field gradient resolution is
roughly equal to the lift height). Furthermore, parameters such as tip
velocity were chosen such that striking of the tip on the surface was
avoided.
A commercial MFM tip was used, the details of the tip
are: force constant: $1-5$N/m; resonant frequency: $50 - 100$~kHz;
nominal tip radius: $35$~nm; magnetic moment:
$1\times10^{-16}$~Am$^2$; magnetic/reflective coating: CoCr.
For the resistance measurements an AC LR700 bridge (four-terminal
sensing) was used with a current peak amplitude of 0.3~mA.

%\onecolumngrid
\begin{figure}
\begin{center}
\includegraphics[width=1\columnwidth]{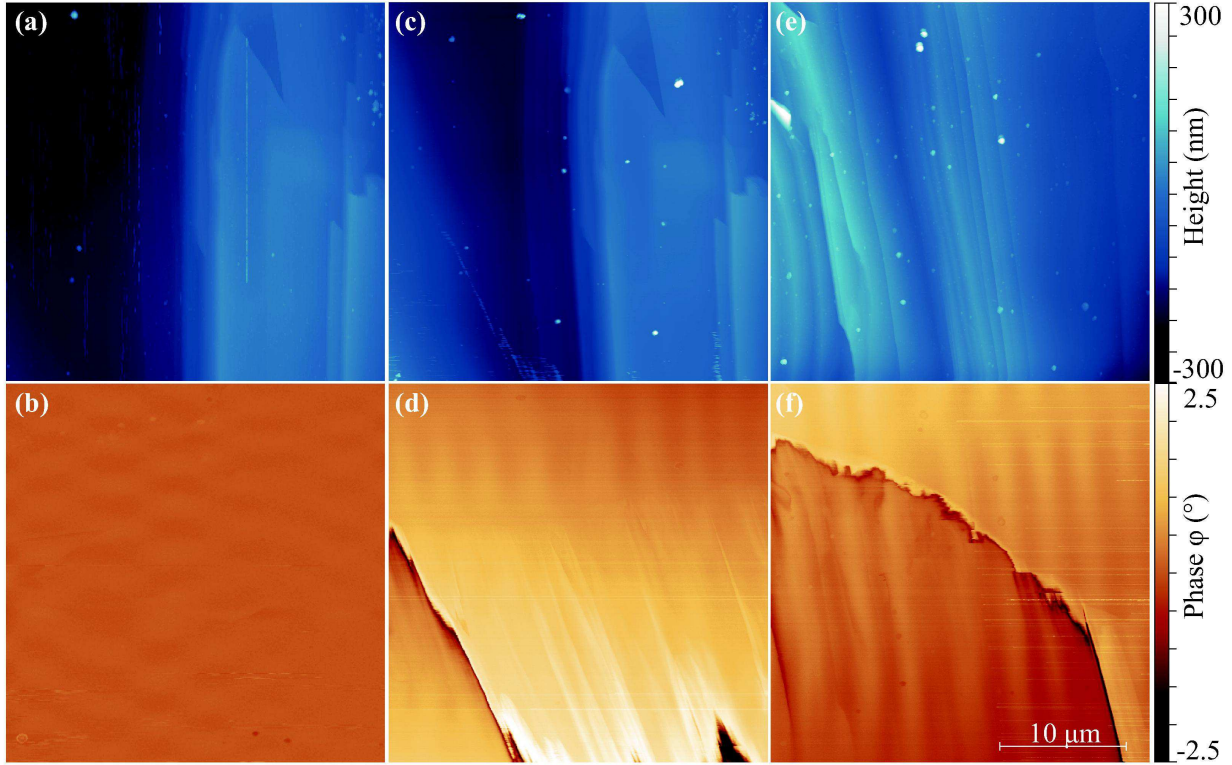}
\caption{Topography ((a),(c),(e)) and phase ((b),(d),(f)) for the
  sample in the virgin state ((a),(b)) and after application of a
  magnetic field ((c)--(f)). The phase shows a signal only after
  application of a magnetic field, independent of
  topography. ((a)--(d)) are results measured in the same region.} \label{fig:fig1}
\end{center}
\end{figure}
%\twocolumngrid

In order to
carry out the temperature dependent MFM measurements from 290~K to 400~K,
a special Cu-plate sample holder was prepared with a heater and a thermometer at
the bottom area. The Cu-plate as well as the sample were
connected to ground.  The substrate with the sample was fixed using
varnish at the upper part of the Cu-plate.  The temperature was
increased such that there was no overshoot, and kept at constant value
throughout the measurements. After the measurements at the maximum
temperature were done, the sample was cooled  to room
temperature without any application of a magnetic field.
No phase shift has been detected in this state. Furthermore, a new tip
was installed and the sample was scanned again with the same
result. Therefore, we rule out the possibility of ferromagnetic order
in the sample.

A calibration of the MFM tip response has been done introducing an
electrical current on Au loops of ring geometry of $\approx1$~$\mu$m
width and $10.5$~$\mu$m diameter patterned by electron beam
lithography. The dependence of the phase signal on current and scan
height were monitored, in order to obtain the effective magnetic
moment and tip-dipole distance~\cite{KC97,WG92}. This can be used to
give only a rough estimation of the magnetic field, as the magnetic
decay length of the sample is unknown. The high temperature measurements have been
done using the same current Au-loop to test the response of the MFM tip.

The very first MFM measurements were done with the sample in the
virgin state. For that the sample was heated to 390~K and cooled down
at zero applied field to 293~K. \figurename~\ref{fig:fig1}(a) and (b)
show the topography and the phase in the maximal scan area
$30 \times 30~\mu$m$^2$ of the sample surface in our device.  The phase
signal, see \figurename~\ref{fig:fig1}(b), does not show any
peculiarity, but a nearly  constant phase value. Different MFM scans of the
sample in this virgin state in a larger area of
$\sim 500 \times 500$~$\mu$m$^2$ provided similar results,
independently of the sample surface topography. Within experimental
resolution, this result indicates the absence of surface stray fields
and rules out the possible existence of magnetic order with magnetic
domains or stray fields at certain topographic peculiarities of the
sample. After applying a magnetic field of $\sim 0.03~$T amplitude and
normal to the sample main area, using a permanent magnet, the
topography and the phase were measured in the same area and shown in
\figurename~\ref{fig:fig1}(c,d).  Whereas the topography remains the
same, as expected, there is a clear feature in the phase, which
indicates the existence of a current path as will become clear below.
\figurename~\ref{fig:fig1}(e,f) show the topology and the phase
signal after application of the magnetic field of another
$30 \times 30~\mu$m$^2$ area of the sample surface.  One can realize
that there is no relationship between the topology and the phase
signal \cite{artefact}.

\begin{figure}
\begin{center}
\includegraphics[width=1\columnwidth]{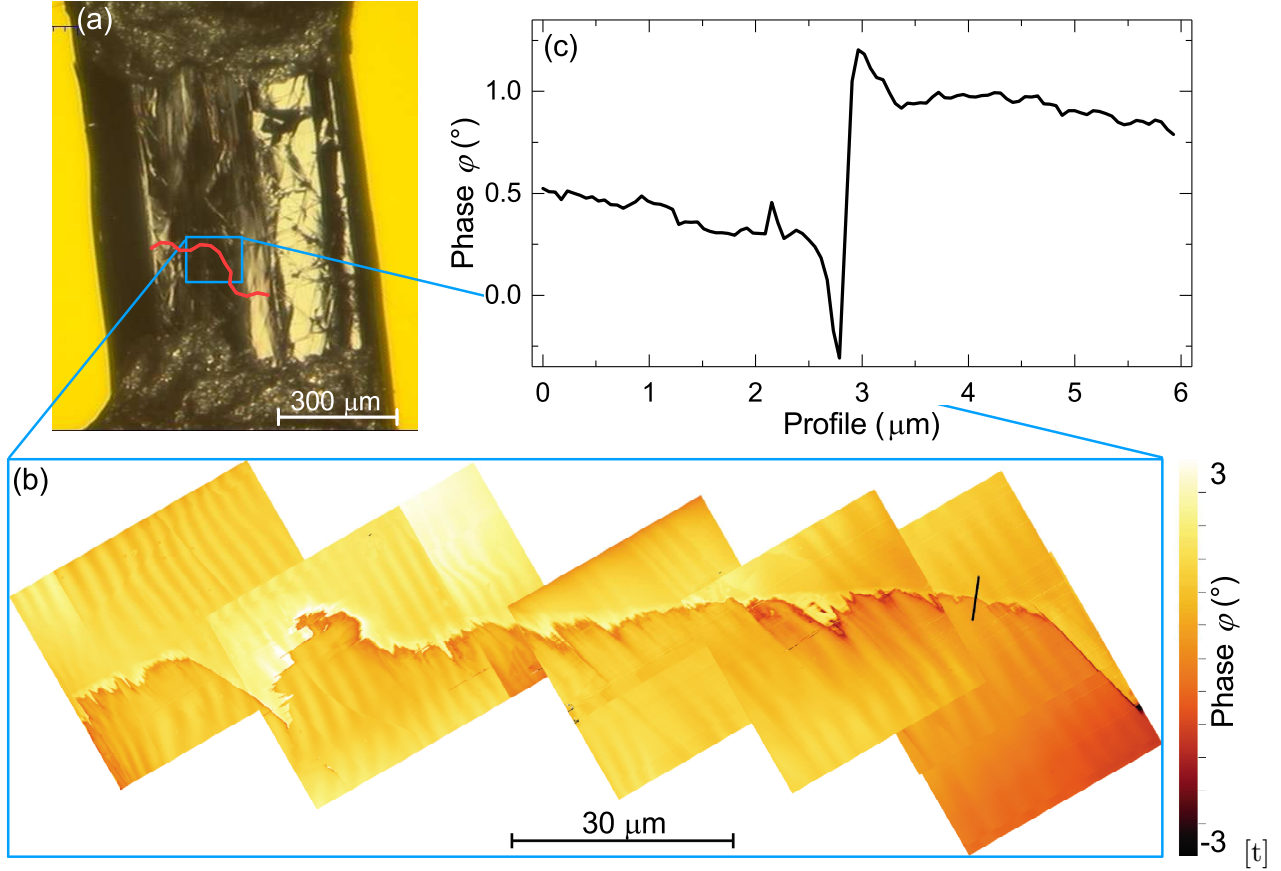}[t]
\caption{In (a) an optical picture of the sample is shown, measurements
  where carried out in the flat regions, following the current path
  indicated as red line. In (b) a composition of MFM measurements can be seen,
  the straight black line shows the position of the phase profile presented in
  (c).} \label{fig:fig2}
\end{center}
\end{figure}

\figurename~\ref{fig:fig2}(a) shows an optical photo of the sample
with its two voltage electrodes at the top and bottom. The red line in the
middle indicates the position of the current path identified by the
MFM measurements. \figurename~\ref{fig:fig2}(b) shows the phase result
in a small portion of this path. We note that the localization and the
measurement of this path took about 6 weeks of continues measurements
(each picture with high resolution and of $30 \times 30~\mu$m$^3$
area takes about $1/2$ day). A line profile of the phase as a function
of position obtained scanning perpendicular through the path
(indicated by the straight black line at the right of
\figurename~\ref{fig:fig2}(b)) is given in
\figurename~\ref{fig:fig2}(c). This phase signal around the edge
region of the phase feature observed at remanence, after applying a
magnetic field, indicates the existence of a current line\cite{kir10}. Moreover, the difference
in the phase between the right and left region from the current line
indicates the existence of a magnetic field of the order of
$\approx3$~mT, following the calibration of the same MFM tip done with Au
current loops.

The finite magnetic field enclosed by a persistent current line implies the
existence of a current loop in the sample. The reason why we could not
measure it completely after 6 weeks of measurements is related to the
cragged topography of the sample at the two edges of the red path seen
in \figurename~\ref{fig:fig2}(a). Therefore, we decided to measure the temperature
dependence of the phase of the current line and compare it with the
electrical resistance results obtained in the same sample as a
function of temperature between 290~K and 390~K.
%\begin{widetext}
%\onecolumngrid
\begin{figure}
\begin{center}
\includegraphics[width=1.\textwidth]{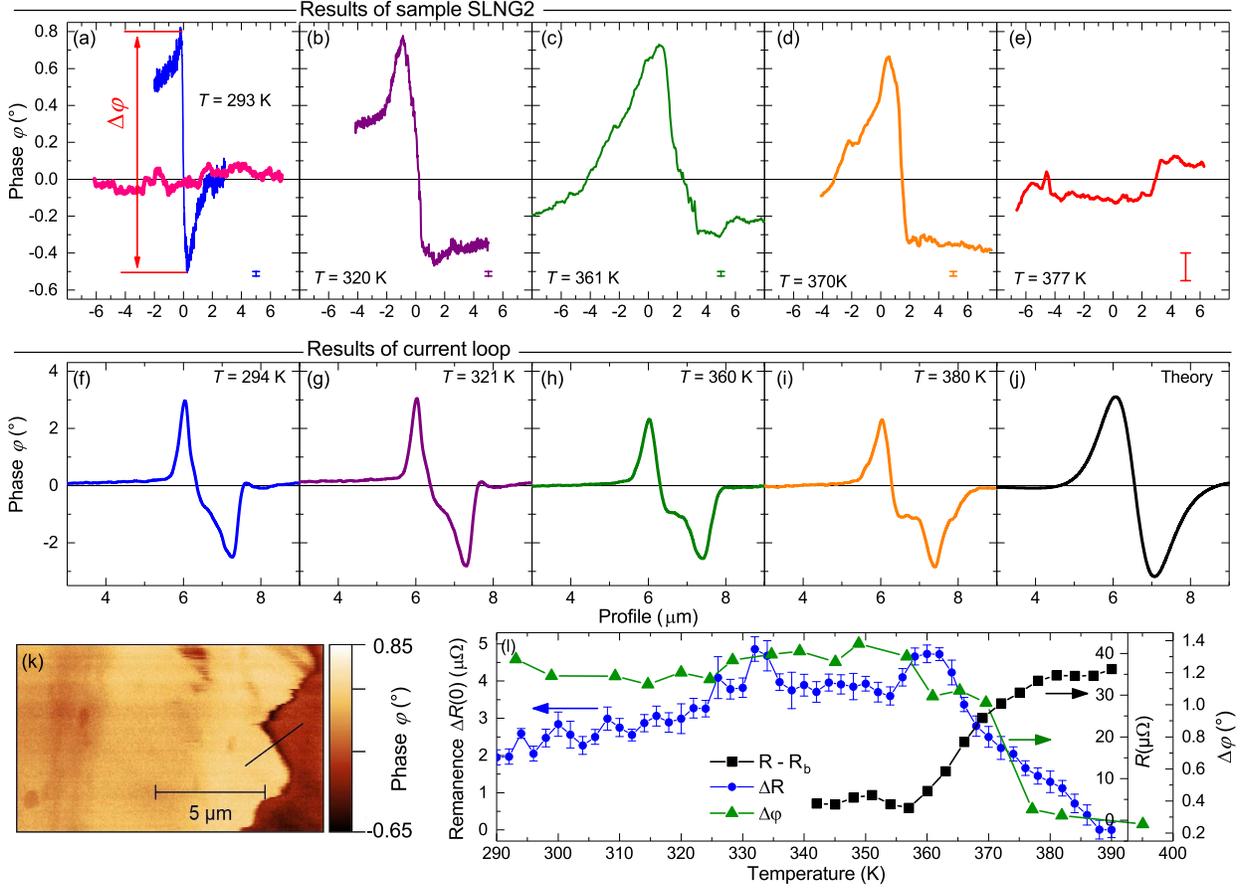}
\caption{ Line scans of the field gradient at different temperatures
  at the edges of the trapped flux region ((a)--(e)) and for a current
  loop of Au of ring geometry of  $\approx 1~\mu$m
  width prepared by electron lithography
  ((f)--(i)).
The bars represent the standard deviation of the
phase at the given temperature.  A theoretical phase shift is shown in (j), obtained
  from  a simulated current line (zero width) loop of  ring geometry. The phase
  image  (k)
  shows the region in the sample surface where the temperature
  dependent measurements were performed. The black line indicates the
  position of the line scans. The well defined difference in phase between the two regions
  separated by the current line is
  because we selected  a U-turn shaped current line (not shown in
  the picture). In Fig.(l), the phase difference $\Delta\varphi$ (see
  (a)), the resistance $R$ (after a linear background subtraction)
  and the remanence $\Delta R(0)$ are shown as
  function of temperature.} \label{fig:fig3}
\end{center}
\end{figure}
%\twocolumngrid
%\end{widetext}
Figures~\ref{fig:fig3}(a)--(e) show the phase line scan through
the black straight line shown in the phase MFM picture in
\figurename~\ref{fig:fig3}(k) at different constant temperatures of
the sample.  Defining $\Delta \varphi$ as indicated in
\figurename~\ref{fig:fig3}(a), which is proportional to the current
amplitude, we plot it as function of temperature in
\figurename~\ref{fig:fig3}(l) (green triangles).  The temperature, at
which this amplitude vanishes, agrees very well with the electrical
resistance measurement as function of temperature at zero field, as
well as with the remanence of the resistance
$\Delta R(0) = R_B(0) - R_0(0)$.  This electrical resistance remanence
$\Delta R(0)$ is defined~\cite{pre16} as the difference between the
resistance measured at zero field after applying a field of 0.03~T
normal to the main surface of the sample $R_B(0)$, and the measured resistance
$R_0(0)$ of the sample in the virgin state obtained after zero-field
cooling from  390~K. The three curves in
\figurename~\ref{fig:fig3}(l) indicate a critical temperature of
$T_c \sim 370~$K with a transition width $\lesssim 30$~K. After
crossing $T_c$ from below and after ZFC to 293~K, the MFM measurement
of the sample does not show any feature at the same position (see magenta
line in \figurename~\ref{fig:fig3}(a)). This vanishing of the remanence
at the same transition temperature obtained by the resistance
measurements, clearly indicates that the origin of the current line is
intimately related to the phenomena we measured from the electrical
resistance.

The fact that this current line obtained from the MFM phase signal
remains for several weeks without decreasing its amplitude within
experimental resolution indicates clearly the existence of a permanent
current, that originates the magnetic field that influences the
electrical resistance. This is a direct proof for the existence of
superconductivity in certain regions of the graphite sample, up to a
critical temperature that lies above room temperature for the
measured sample. As shown in previous publications, the
magnetization~\cite{sch12} as well as the electrical
resistance~\cite{pre16} reveal a flux creep phenomenon with a
logarithmic time dependence. Those results suggest that the
current line measured by the MFM phase should also show some kind of time
dependence. Measuring the evolution of the MFM phase line with
time we identified a shift in its position compatible with the decrease
with time of the field enclosed area due to flux creep, see the video included in the
supplementary information.

The meandering structure of the current path shown by the MFM phase
line, see \figurename~\ref{fig:fig2}(b), is similar to the one
observed in high-temperature superconducting oxides in
remanence~\cite{vla97,vla98}. In that case a modified Bean model
based on a finite lower critical field $H_{c1}$ and vortex pinning was
used to understand the origin of the current line known as the
Meissner hole (for a review see~\cite{joo02}).  In the case of our
graphite sample, however, from the simple Ginzburg-Landau relation for
$H_{c1} \propto 1/\lambda_{\perp}^2$,  taking the effective
penetration depth as the one derived by Pearl \cite{pea64} for very thin films
$\lambda_{\perp} = 2 \lambda_L^2/d \gg \lambda_L$ ($d$ is the thickness
of the superconducting interface $\sim$ distance between graphene
planes) one expects a negligible $H_{c1}$. Moreover, due to the
expected huge penetration depth, the pinning of pancake vortices at
the interfaces should be rather negligible at such high temperatures.
In contrast, the maximum in the remanence measured by the resistance is just below
the transition (blue points in \figurename~\ref{fig:fig3}(l)). This fact
suggests that the magnetic field at remanence is produced by
macroscopic (or mesoscopic) current loops, which originate fluxons. These
pinned fluxons are the origin for the remanent state of the magnetic field and
the irreversible behavior observed in  the electrical resistance.

In conclusion,  through MFM measurements done on a natural
graphite sample that shows a transition in the electrical
resistance and its remanence at $T_c \simeq 370~$K we could
localize a current line as the origin for the trapped flux. This
current remains for several weeks basically unchanged but it shows
creep. The  current line vanishes irreversibly at the same
temperature as the electrical resistance  shows a transition. Our
results indicate that MFM as well as other scanning magnetic
imaging techniques can be used to identify the regions of graphite
samples were superconductivity is localized. This will undoubtedly
help to further characterize the superconducting interfaces and/or
other regions of interest in graphite, paving the way for their
future device implementations.

\begin{acknowledgments}  We thank Henning Beth for providing us with the natural graphite sample from
Sri Lanka. C.E.P. gratefully acknowledges
 the support provided by The Brazilian National Council for the Improvement of
 Higher Education (CAPES). M.S. and J.B-Q. are supported by the DFG
 collaboration project SFB762.
\end{acknowledgments}

%\bibliography{magnetic_carbon_MS}

\begin{thebibliography}{19}
\bibitem [{art()}]{artefact}%
  \BibitemOpen
  \href@noop {} {}\bibinfo {note} {There are however regions with a periodic
  modulation of the phase, see Fig. 2(b), which remain unaffected by the sample
  state and are due to an artefact of the used MFM.}\BibitemShut {Stop}%
\end{thebibliography}
%\clearpage


\end{document}